# A basic problem in the correlations between statistics and thermodynamics

Congjie Ou[1,2], Zhifu Huang[1] and Jincan Chen[1,*]

[1]Department of Physics and Institute of Theoretical Physics and Astrophysics, Xiamen University, Xiamen 361005, People's Republic of China

[2]College of Information Science and Engineering, Huaqiao University, Quanzhou 362021, People's Republic of China.

**Abstract:** By using a fonctionelle of probability distributions, several different statistical physics including extensive and nonextensive statistics are unified in a general method. The essential equivalence between the MaxEnt process of the most probable probility distribution in these statistics and the famous thermodynamical relation $dU = TdS$ is strictly proved without any additional assumption. Moreover, it is expounded that all the conclusions of these different statistics can be directly derived from the equivalent relation.

Email: jcchen@xmu.edu.cn

The probability distribution function is very important in the statistical physics. It's a bridge between large numbers of particle movement principles in microscopic view and the thermodynamical phenomena in macroscopic view, so the hypothesis proposed by Boltzmann and the thermodynamical physics can be unified into a whole. The variation of the probability distribution always affects the thermodynamic properties of a system simultaneously. The probability distribution function determines the number of particles in each energy state [1-3] and the variations of particle number in the microstates lead to energy exchanges between the system and the heat bath. How are they connected by basic thermodynamic principles? This is a very fundamental problem. Although the important problem was recently proposed by Plsatino and Curado [4], it has not been completely solved. In this Letter, the connections between the different statistical physics including extensive [1-3] and nonextensive [4-15] statistics and thermodynamics will be investigated by a unified method and a very general conclusion will be obtained.

For different statistical physical methods, the forms of the entropy are very different each other. Here a fonctionelle $f(p_i)$ of the probability distribution $\{p_i\}$ $(i=1,2,3...W)$ can be used to unify the different forms of entropy,

$$S = k\sum_{i=1}^{W} p_i^{q^\delta} f(p_i), \qquad (1)$$

where $k$ is the Boltzmann constant, $p_i$ is the probability of the state $i$ among $W$ possible ones that are accessible to the calculation, $q$ is a real number or called a nonextensive parameter, and $\delta$ is an information criterion. The entropy or the information function of a system is based on the sum of the total microstates. The principles of quantum mechanics tell us that all the total energy states of a system can be derived from the Schrödinger Equation (SE). If the SE of a system can be totally solved, it means that all of the $W$ microstates are available [1, 3]. In this case the information of the system is complete, the information criteion can be taken as $\delta = 0$. On the contrary, if there exist some complex interactions in a system, the solutions of the SE can't be exactly solved, so that one can only get $W$ microstates of the system through some approximations. In this case $W$ may be more or less than the number of real energy states of the system [7], and then the information of the system is incomplete and taken as $\delta = 1$. For the sake of convenience, $\sum_{i=1}^{w}$ is replaced by $\sum_{i}$ below.

As described in literature and textbooks, the normalizations of the probability distribution of Tsallis' statistics and extensive statistics are simultaneously given by [1, 2, 5, 6]

$$\sum_i p_i = 1, \tag{2}$$

because the complex long-range interaction and/or long-duration memory in Tsallis' statistics may be embodied in the nonextensive paramenter $q$. While in the Incomplete Statistics (IS), the normalization of the probability distribution of IS is different from that of the extensive statistics and given by [7]

$$\sum_i p_i^q = 1, \tag{3}$$

where $q$ is a nonnegative real number and correlated with the Hausdorff dimensions in incomplete fractal phase space [8]. Equations (2) and (3) can be written as a unified expression, i.e.,

$$\sum_i p_i^{q^\delta} = 1, \tag{4}$$

where the nonextensive parameter extends its region to $q \in R$. On the other hand, the observable quantities of a physical system can be written as the weighted expectation of the microscopic states, i.e.,

$$\langle O \rangle = \sum_i O_i g(p_i), \tag{5}$$

where $g(p_i)$ is another fonctionelle of the probability distribution function. The concrete forms of $g(p_i)$ are decided by different statistical methods. Some representational examples will be illustrated below. Based on Eq. (5), the internal energy of a system can be expressed as

$$U = \sum_i \varepsilon_i g(p_i). \tag{6}$$

Now, we focus our interest on the discussion of a canonical ensemble since it's the most important base of all kinds of statistical physics. The maximum entropy (MaxEnt) method in a canonical ensemble yields the probability distribution of the system in equilibrium. The thermal equilibrium between the system and the heat bath makes the entropy of the system reaching its maximum and the sum of energy exchanges is equal to zero. The normalization of the probability distribution must be considered simultaneously, so the Lagrange method can be written as

$$\delta_{\{p_i\}} \left[ \frac{S}{k} - \beta U - \alpha \sum_i p_i^{q^\delta} \right] = 0. \tag{7}$$

Substituting Eqs. (1) and (4) into Eq. (7), one can get

$$\sum_i \left[ f(p_i) + \frac{p_i f'(p_i)}{q^\delta} - \frac{\beta U'}{q^\delta p_i^{q^\delta - 1}} - \alpha \right] q^\delta p_i^{q^\delta - 1} dp_i = 0, \tag{8}$$

where ' means the derivative of a function with respect to $p_i$. Using Eq. (8) and the differential of Eq. (4),

$$\sum_i q^\delta p_i^{q^\delta - 1} dp_i = 0, \tag{9}$$

we have

$$f(p_i) + \frac{p_i f'(p_i)}{q^\delta} - \frac{\beta U'}{q^\delta p_i^{q^\delta - 1}} - \alpha = 0. \tag{10}$$

It is well known that there exists a relation between the physical temperature of the system and the Lagrange multiplier $\beta$ which is correlated with internal energy [9], i.e.,

$$\beta = \frac{1}{kT}. \tag{11}$$

If the microscopic energy states $\varepsilon_i$ ($i=1,2,3...W$) are fixed, it means that the system has no work exchange with the circumstance in macroscopic view and the variations of the probability distribution of each microstates are only aroused by the particles transmitting among energy states. These transitions are stochastic and reversible. According to the first law of thermodynamics, the physical meaning of these transitions is an infinitesimal reversible heat exchange between the system and the heat bath, so the variation of internal energy of the system satisfies a reversible Clausius relation and is given by

$$dU = dQ = TdS. \tag{12}$$

Substituting Eq. (1) into Eq. (12) one can obtain

$$\sum_i \left[ \frac{U'}{q^\delta p_i^{q^\delta-1}} - kT\left( f(p_i) + \frac{p_i f'(p_i)}{q^\delta} \right) \right] q^\delta p_i^{q^\delta-1} dp_i = 0. \tag{13}$$

Combining Eq. (13) with Eq. (9) yields

$$F_i = F_j = F, \quad (i,j=1,2,3,...,W) \tag{14a}$$

or

$$\frac{U'}{q^\delta p_i^{q^\delta-1}} - kT\left( f(p_i) + \frac{p_i f'(p_i)}{q^\delta} \right) = F. \tag{14b}$$

Comparing Eq. (14) with Eq. (10), one can easily find that they are the same if $\alpha = -\beta F$. It shows clearly that the essential equivalence between the MaxEnt method and the famous thermodynamical relation $dU = TdS$ has been illustrated by mathematical principles.

Below some typical examples would illuminate some concrete applications of this equivalence.

For the cases of $\delta = 0$, the information of systems is complete and Eq. (14) may be simplified as

$$f(p_i) + p_i f'(p_i) - \beta U' - \alpha = 0. \tag{15}$$

Starting from Eq. (15), one can easily derive the probability distributions of not only classical Boltzmann-Gibbs statistics [1, 2] but also Tsallis' statistics [5, 6, 10] and the exponential entropic form [11]. For example, in Tsallis' statistics with the normalized energy constraint, $f(p_i)$, $f'(p_i)$ and $U'$ can be, respectively, expressed as [6, 12]

$$f(p_i) = \frac{1 - p_i^{q-1}}{q-1}, \quad f'(p_i) = -p_i^{q-2} \tag{16}$$

and

$$U' = \frac{q p_i^{q-1}}{\sum_j p_j^q}(\varepsilon_i - \sum_k p_k^q \varepsilon_k / \sum_l p_l^q) = \frac{q p_i^{q-1}}{\sum_j p_j^q}(\varepsilon_i - U). \tag{17}$$

Substituting Eqs. (16) and (17) into Eq. (15) and using Eq. (2) yields

$$p_i = \frac{1}{Z}\left[1 - (1-q)\beta(\varepsilon_i - U)/\sum_j p_j^q\right]^{1/(1-q)}, \tag{18}$$

where

$$Z = \sum_i \left[1 - (1-q)\beta(\varepsilon_i - U)/\sum_j p_j^q\right]^{1/(1-q)} \tag{19}$$

is the partition function of the system [10]. Equation (18) is just the normalized probability distribution in the Tsallis-Mendes-Plastino (TMP) fashion [6]. It is significant to note that the additional assumption of the uniform energy-shift [4] is unnecessary for the above derivative process. When $U' = \varepsilon_i g'(p_i)$ is true, Eq. (15) can be simplified as

$$f(p_i) + p_i f'(p_i) - \beta \varepsilon_i g'(p_i) - \alpha = 0, \tag{20}$$

which is just Eq. (9) in Ref. [4]. For example, in Tsallis' statistics with the second energy constraint (unnormalized constraint), $U'$ is determined by [4]

$$U' = \varepsilon_i g'(p_i) = q \varepsilon_i p_i^{q-1}. \tag{21}$$

Substituting Eqs. (16) and (21) into Eq. (20) and using Eq. (2), one has

$$p_i = \frac{1}{Z}[1 - (1-q)\beta \varepsilon_i]^{1/(1-q)}, \tag{22}$$

where

$$Z = \sum_i [1-(1-q)\beta\varepsilon_i]^{1/(1-q)} \qquad (23)$$

is the partition function of the system [6, 10]. Similarly, Eq. (15) or (20) may be used to calculate the probability distribution of the classical Boltzmann-Gibbs statistics, Tsallis' statistics with the first energy constraint and the exponential entropic form [11].

For the cases of $\delta = 1$, the information of systems is unclear or incomplete to some extent [7, 8, 13, 14] and Eq. (15) can be expressed as

$$f(p_i) + \frac{p_i f'(p_i)}{q} - \frac{\beta U'}{q p_i^{q-1}} - \alpha = 0. \qquad (24)$$

In IS, $f(p_i)$, $f'(p_i)$ and $U'$ can be, respectively, expressed as [7]

$$f(p_i) = \frac{p_i^{1-q}-1}{q-1}, \qquad f'(p_i) = -p_i^{-q} \qquad (25)$$

and

$$U' = \varepsilon_i g'(p_i) = q\varepsilon_i p_i^{q-1}. \qquad (26)$$

Substituting Eqs. (25) and (26) into Eq. (24) and using Eq. (3) yields the probability distribution of IS as [7,14]

$$p_i = \frac{1}{Z}[1-(1-q)\beta^*\varepsilon_i]^{1/(1-q)}, \qquad (27)$$

where $Z = \left\{\sum_i [1-(1-q)\beta^*\varepsilon_i]^{q/(1-q)}\right\}^{1/q}$ and $\beta^* = \frac{\beta}{1-(q-1)\beta F}$.

To sum up, a general principle between the different statistical methods including extensive and nonextensive statistics and thermodynamics is discussed. For a canonical ensemble, the effect of infinitesimal disturbance of the probability distribution is equivalent with a reversible heat exchange between the system and the heat bath in a process without work in macroscopic view. The MaxEnt method and the famous thermodynamical relation $dU = TdS$ are essentially the same. This equivalence can be described by a general mathematical expression. All of the different kinds of statistics obey this basic principle which has been strictly proved in this Letter.The conclusion obtained here conforms to Abe's standpoint [15], i.e., statistical mechanics may be modified but thermodynamics should remain unchanged.

**Acknowledgements**

This work has been supported by the Research Foundation of Ministry of Education, People's Republic of China and the Science Research Fund (No. 07BS105), Huaqiao University, People's Republic of China.